\documentclass[12pt]{article}
\usepackage{graphicx}
\usepackage[cp1251]{inputenc}
\usepackage{epsfig}
\usepackage[english]{babel}

\date{}
\def\b{\begin{equation}}
\def\e{\end{equation}}
\def\bee{\begin{enumerate}}
\def\eee{\end{enumerate}}
\def\be{\begin{vmatrix}}
\def\ee{\end{vmatrix}}

\begin{document}
\setcounter{page}{1}
\bibliographystyle{unsrt2}
\pagestyle{plain}

\title{\bf{Effects of spin-orbit coupling and magnetic field on electronic properties of
Germanene structure}}
\author{Farshad Azizi and Hamed Rezania\thanks{Corresponding author.
Tel./fax: +98 831 427 4569., Tel: +98 831 427 4569.
E-mail: rezania.hamed@gmail.com}}
\maketitle{\centerline{Department of Physics, Razi University,
Kermanshah, Iran}
\begin{abstract}
In this paper, we present a Kane-Mele model in the presence of
magnetic field and next nearest neighbors hopping amplitudes for investigations the electronic and
optical properties of monolayer Germanene.
Specially, we address the dynamical conductivity of the structure
as a function of photon frequency
and in the presence of magnetic field and spin-orbit coupling
 at finite temperature. Using linear response theory and
Green's function approach, the frequency dependence of optical conductivity has been obtained
 in the
context of Kane-Mele model Hamiltonian.
Our results show a finite Drude response at low frequency
at non zero value for magnetic field in the presence of spin-orbit coupling.
However Drude weight gets remarkable amount in the presence of electron doping.
The thermal conductivity and specific heat increase with increasing the temperature at low amounts
of temperature due to the increasing of thermal energy of charge carriers and excitation of them
to the conduction bands.
The results for Seebeck coefficient show the sign of thermopower is negative in the
presence of spin-orbit coupling. Also we have studied the temperature dependence of
electrical conductivity of Germanene monolayer due to both spin orbit coupling and magnetic field
factors in details.
\end{abstract}
\vspace{0.5cm} {\it \emph{Keywords}}: Germanene; Green's function; Optical absorption
\section{Introduction}

Graphene as a one-atom-thick layer of graphite, attracts a lot of attention of both theoreticians
and experimentalists since it's fabrication\cite{novoselov}. Initially studies of graphene were limited to
realm of theory where the low energy linear dispersion and chiral nature of the honeycomb carbon lattice
were shown to result from a simple the nearest neighbor hopping tight binding hamiltonian which
at low energy maps on to a Dirac Hamiltonian for massless
 fermions with Fermi velocity $v_{F}$.
Graphene layer with a zero band gap energy exhibit some interesting electronic properties
compared to materials with a non-zero energy gap and they have intriguing physical properties and
numerous potential practical applications in spintronics, electronics,
optics and sensors\cite{wang}. The density functional theory calculations shows the optical
absorption takes place in a wide range spectra which leads to the applications of graphene in
electro-optical devices\cite{mates,falkovsky}.
The optical properties of graphene are of considerable importance
for technological applications, all variants of graphene are also of
potential interest and should be examined. The dynamical
conductivity of graphene has been extensively studied
theoretically\cite{ando} and experiments have largely verified the
expected behavior\cite{nair}.

Recently, the hybrid systems consisting of graphene and various two-dimensional
materials have been studied extensively both experimentally and theoretically
\cite{liu,chang1,chang2}.
Also, the use of 2D materials could be advantageous for a wide range applications in
nanotechnology \cite{dean,novo} and memory technology \cite{bertol}.
While the research interest in graphene-based superlattices is growing rapidly,
 people have started
to question whether the graphene could be replaced by its close relatives, such as 2 dimensional
hexagonal crystal of Germanene. This material is zero gap semiconductors with massless fermion
charge carriers since their $\pi$ and $\pi^{*}$ bands are also linear at
 the Fermi level\cite{akturk}.
Germanene as counterpart of graphene, is predicted to have a geometry with low-buckled
honeycomb structure for its most stable structures in contrast to the graphene monolayer
\cite{akturk,ccliu}. Such small buckling as vertical distance between two planes of
atoms for Germanene comes from the mixing of $sp^{2}$ and
$sp^{3}$ hybridization\cite{padil,chowd}. The electronic structure of Germanene behaves linear
dispersion around K and K' points of the first Brillouin zone. However {\it ab initio} calculations
demonstrated that spin-orbit coupling in Germanene causes to small band gap opening at the Dirac
point and consequently the Germanene has massive Dirac fermions\cite{ccliu, kaloni}.
Also the band gap due to the spin orbit coupling in Germanene is more considerable
 rather than that in Graphene\cite{liu1}.
The intrinsic carrier mobility of Germanene is higher than graphene\cite{ye}.
The different dopants within the Germanene layer leads to the sizable band gap opening at the
Dirac point and the electronic properties of this material are affected by that\cite{monshi,sun}.
There are some studies concerning band gap opening due to different external dopant.
Electronic optical, properties and widening band gap of graphene with Ge doping have been
investigated\cite{ould}.
The effects of external dopant on optoelectronics of graphene have been studied in
theoretical work\cite{Bonaccorso}.
Also the chemical reactivity and bandgap opening of graphene doped with Gallium,
Germanium, Arsenic,
 and Selenium atoms have been addressed in the other theoretical work\cite{denis}.
Band gap opening in dual-doped monolayer graphene has been studied by the other
theoretical group\cite{denis1}.

In a theoretical work, the structural and electronic properties of superlattices made
with alternate stacking of Germanene layer are systematically investigated by using a density
functional theory with the van der Waals correction\cite{xia}.
 It was found that intrinsic spin-orbit coupling due to a perpendicular
 electric field or interaction
with a substrate plays an important role on the topological properties of graphene
like structures. It was predicted that spin orbit coupling and exchange field together open a
nontrivial bulk gap in graphene like structures leading to the quantum spin hall effect
\cite{qiao,tse}. The topological phase transitions in the 2D crystals can be understood
based on intrinsic spin orbit coupling which arises due to
 perpendicular electric field or interaction with a substrate.
A simple model introduced by Kane and Mele\cite{kane} has been applied to describe topological insulators.
Such model consists of a hopping and an intrinsic spin-orbit term on the graphene like structures.
The Kane-Mele model essentially includes two copies with different sign for up and down spins of
a model introduced earlier by Haldane\cite{haldan}.
As the first microscopic model for topological insulators, the Kane-Mele model was originally proposed
to describe the quantum spin Hall effet in graphene\cite{kane}. Subsequent band structure
calculations showed, however, that the spin orbit gap in graphene is so small\cite{min,yao} that the quantum spin
Hall effect in graphene like structures is beyond experimental relevance.

Moreover it has been shown that in-plane magnetic field induces honeycomb structures magneto
resistance which is negative for intrinsic gapless graphene while for extrinsic gapless
graphene, magneto-resistance is a positive value for fields lower than the critical magnetic
field and negative above the critical magnetic\cite{hwang}.
Microwave magneto transport in doped graphene
is an open problem\cite{mani}.
The theoretical investigations have been done to study the optical and thermal properties of
Germanene\cite{mohan,john}. By finding the photon dispersions of Germanene, it was found that the
Germanene is stable by applying and increasing the strain up to 16 percent and in
this range the Dirac cone shifts towards higher energy\cite{kaloni1}. The optical properties of
Germanene layer show that the optical absorption is shifted from ultraviolet to infrared from
graphene to Ge\cite{john} and this material has remarkable light absorption and presents optical anisotropy\cite{john}.

The purpose of this paper is to provide a Kane Mele model including intrinsic spin-orbit interaction
for studying the electrical and thermal properties of Germanene layer in the presence of
magnetic field perpendicular to the plane.
Using the suitable hopping integral and on site parameter values, the band dispersion of
electrons has been calculated.
For calculating the transport coefficients we have exploited the
linear response theory in the context of Kubo formula. Based on Green's function approach, we have obtained
the density of states, thermal conductivity and Seebeck coefficient of Germanene monolayer.
Moreover the photon frequency behavior of optical absorption has been studied for different
values of spin-orbit coupling strength.
 The effects of electron doping and spin orbit coupling on the frequency behavior of
 optical absorption of Germanene layer
 have been studied. Also we discuss and analyze to
show how spin orbit coupling and longitudinal magnetic field affect the temperature behavior of thermal properties of Germanene.
 \section{Model Hamiltonian and formalism}
The crystal structure of Germanene has been shown in Fig.(1).
The unit cell of Germanene structure is similar to Graphene layer and this honeycomb lattice
 depicted in Fig.(2). The primitive unit cell vectors of honeycomb lattice
have been shown by ${\bf a}_{1}$ and ${\bf a}_{2}$.
However Germanene is buckled due to sp2-sp3 combination of
 hybridization, the density functional theory calculations
demonstrate the hopping amplitudes are
dominant for electrons in orbital $p_{z}$\cite{chegel}.
In other words $\pi$ electrons gives the major
contribution to the band structure of electrons.
The mentioned reference has been cited in revised version.

In the presence of longitudinal magnetic field,
the Kane-Mele model\cite{kane} ($H$)
for Germanene structure includes the tight binding model ($H^{TB}$), the intrinsic spin-orbit coupling ($H^{ISOC}$)
and the Zeeman term ($H^{Zeeman}$) due to the coupling of spin degrees of freedom of electrons with external longitudinal magnetic field
 $B$
\begin{eqnarray}
 H&=&H^{TB}+H^{ISOC}+H^{Zeeman}.
 \label{e0.5}
\end{eqnarray}
The tight binding part of model Hamiltonian
consists of three parts; nearest
neighbor hopping, next nearest neighbor (2NN) hopping and
 next next nearest neighbor (3NN) hopping terms. The tight binding part, the spin orbit coupling term
and the Zeeman part of the model Hamiltonian on the honeycomb lattice are given by
\begin{eqnarray}
 H^{TB}&=&-t\sum_{i,\Delta,\sigma}\Big(a^{\sigma\dag}_{j=i+\Delta}b^{\sigma}_{i}+h.c.\Big)
-t'\sum_{i,\Delta',\sigma}\Big(a^{\sigma\dag}_{j=i+\Delta'}a^{\sigma}_{i}+b^{\sigma\dag}_{i+\Delta'}b^{\sigma}_{i}\Big)
-t"\sum_{i,\Delta",\sigma}\Big(a^{\sigma\dag}_{j=i+\Delta"}b^{\sigma}_{i}+h.c.\Big)
\nonumber\\&-&\sum_{i,\sigma}
\mu\Big(a^{\dag\sigma}_{i}a^{\sigma}_{i}+b^{\dag\sigma}_{i}
b^{\sigma}_{i}\Big),\nonumber\\
H^{ISOC}&=&
i\lambda\sum_{i,\Delta',\sigma}\sum_{\alpha=A,B}\Big(\nu^{a}_{i+\Delta',i}
a^{\dag\sigma}_{i+\Delta'}\sigma^{z}_{\sigma\sigma'}
a^{\sigma'}_{i}+\nu^{b}_{i+\Delta',i}
b^{\dag\sigma}_{i+\Delta'}\sigma^{z}_{\sigma\sigma'}
b^{\sigma'}_{i}\Big),\nonumber\\
H^{Zeeman}&=&-\sum_{i,\sigma}\sigma g\mu_{B}B\Big(a^{\dag\sigma}_{i}a^{\sigma}_{i}+b^{\dag\sigma}_{i}
b^{\sigma}_{i}\Big).
\label{e1}
\end{eqnarray}
Here $a^{\sigma}_{i}(b^{\sigma}_{i})$ is an annihilation operator of electron with
spin $\sigma$ on sublattice
$A(B)$ in unit cell index $i$. The operators fulfill the fermionic standard
anti commutation relations $\{a^{\sigma}_{i},a^{\sigma'\dag}_{j}\}=
\delta_{ij}\delta_{\sigma\sigma'}$. As usual $t,t',t''$ denote the nearest neighbor, next
nearest neighbor and next next nearest neighbor
 hopping integral amplitudes, respectively. The parameter $\lambda$ introduces the
spin-orbit coupling strength. $\sigma^{z}$ is the third Pauli matrix, and $\nu_{ji}^{a(b)}=\pm 1$
as discussed below.
Based on Fig.(2), ${\bf a}_{1}$ and ${\bf a}_{2}$ are the primitive vectors of unit
cell and the length of them is assumed to be unit.
The symbol $\Delta={\bf 0},\vec{\Delta}_{1},\vec{\Delta}_{2}$
 implies the indexes of
lattice vectors connecting the unit cells including nearest neighbor lattice sites.
The translational vectors  $\vec{\Delta}_{1},\vec{\Delta}_{2}$ connecting
 neighbor unit cells are given by
\begin{eqnarray}
\vec{\Delta}_{1}={\bf i}\frac{\sqrt{3}}{2}+{\bf j}\frac{1}{2}\;\;,\;\;\vec{\Delta}_{2}=
{\bf i}\frac{\sqrt{3}}{2}-{\bf j}\frac{1}{2}.
\label{e4}
\end{eqnarray}
Also index $\Delta'=\vec{\Delta}_{1},\vec{\Delta}_{2},
-\vec{\Delta}_{1},-\vec{\Delta}_{2},{\bf j},-{\bf j}$
implies the characters of lattice
vectors connecting the unit cells including next nearest neighbor lattice sites.
Moreover index $\Delta"=\sqrt{3}{\bf i},{\bf j},-{\bf j}$
denotes the characters of lattice
vectors connecting the unit cells including next next nearest neighbor lattice sites.
We consider the intrinsic spin-orbit term\cite{kane}
 of the KM Hamiltonian in Eq.(\ref{e1}). The expression $\nu_{ji}^{a(b)}$ gives $\pm 1$
depending on the orientation of the sites. A standard definition for $\nu^{\alpha}_{ji}$ in each sublattice $\alpha=A,B$ is $\nu_{ji}^{\alpha}=\Big(
\frac{{\bf d}^{\alpha}_{j}\times {\bf d}^{\alpha}
_{i}}{|{\bf d}^{\alpha}_{j}\times {\bf d}^{\alpha}_{i}|}\Big)_{z}=\pm 1$ where ${\bf d}
^{\alpha}_{j}$ and ${\bf d}
^{\alpha}_{i}$ are the two unit vectors along the nearest neighbor bonds connecting site $i$
to its next-nearest neighbor $j$. Because of two sublattice atoms, the band wave function
$\psi_{n}({\bf k},{\bf r})$ can be expanded in terms of Bloch functions
$\Phi_{\alpha}({\bf k},{\bf r})$. The index $\alpha$ implies two inequivalent sublattice atoms
$A,B$ in the unit cell, ${\bf r}$ denotes the position vector of electron, ${\bf k}$ is
the wave function belonging in the first Brillouin zone of honeycomb structure.
Such band wave function can be written as
\begin{eqnarray}
\psi_{n}({\bf k},{\bf r})=\sum_{\alpha=A,B}C^{n}_{\alpha}({\bf k})
\Phi_{\alpha}({\bf k},{\bf r}),
 \label{e5}
\end{eqnarray}
where $C^{n}_{\alpha}({\bf k})$ is the expansion coefficients and $n=c,v$ refers to condition
and valence band. Also we expand the Bloch wave function in terms of Wannier wave function as
\begin{eqnarray}
\Phi_{\alpha}({\bf k},{\bf r})=\frac{1}{\sqrt{N}}\sum_{{\bf R}_{i}}
e^{i{\bf k}.{\bf R}_{i}}\phi_{\alpha}({\bf r}-{\bf R}_{i}),
 \label{e6}
\end{eqnarray}
so that ${\bf R}_{i}$ implies the position vector of $i$th unit cell in the crystal and
$\phi_{\alpha}$ is the Wannier wave function of electron in the vicinity of
atom in $i$ th unit cell on sublattice index $\alpha$. The small Buckling in Germanene causes
to the considerable value for 2NN and 3NN hopping amplitude. Moreover we have considerable values for overlap parameters of electron
wave functions between 2NN and 3NN atoms. The band structures of electrons of Germanene described by model Hamiltonian in Eq.(\ref{e1}) are
obtained by using the matrix form of Schrodinger as follows
\begin{eqnarray}
{\mathcal H}({\bf k}){\mathcal C}({\bf k})&=&E_{n}({\bf k}){\mathcal S}
({\bf k}){\mathcal C}({\bf k}),\nonumber\\
{\mathcal H}({\bf k})&=&\left(
                              \begin{array}{cc}
                               H_{AA}({\bf k})&   H_{AB}({\bf k}) \\
                               H_{BA}({\bf k}) &  H_{BB}({\bf k}) \\
\end{array}
\right)\;,\;{\mathcal C}({\bf k})=\left(
                              \begin{array}{c}
                               C^{n}_{A}({\bf k})\\
                               C^{n}_{B}({\bf k})\\
\end{array}
\right),\;\;\nonumber\\
{\mathcal S}({\bf k})&=&\left(
                              \begin{array}{cc}
                               S_{AA}({\bf k})&   S_{AB}({\bf k}) \\
                               S_{BA}({\bf k}) &  S_{BB}({\bf k})\\
\end{array}
\right).
 \label{e7}
\end{eqnarray}
Using the Bloch wave functions, i.e. $\Phi_{\alpha}({\bf k})$, the matrix elements of ${\mathcal H}$ and ${\mathcal S}$ are given by
\begin{eqnarray}
 H_{\alpha\beta}({\bf k})=\langle\Phi_{\alpha}({\bf k})|{\mathcal H}|\Phi_{\beta}({\bf k})\rangle\;\;,\;\;
 S_{\alpha\beta}({\bf k})=\langle\Phi_{\alpha}({\bf k})|\Phi_{\beta}({\bf k})\rangle.
 \label{e8}
\end{eqnarray}
The matrix elements of $H_{\alpha\beta}$ and $S_{\alpha\beta}$ are expressed based on hopping amplitude and spin-orbit coupling
between two neighbor atoms on lattice sites and can be expanded in terms of hopping amplitudes $t,t',t"$,
spin orbit coupling $\lambda$ and overlap parameters. The diagonal elements of matrixes ${\mathcal H}$
 in Eq.(\ref{e7}) arise from hopping amplitude of electrons between next nearest neighbor atoms on the same sublattice
 and spin-orbit coupling.
Also the off diagonal matrix elements $H_{AB}, H_{BA}$ raise from hopping amplitude of electrons between
 nearest neighbor atoms and next next nearest neighbor atoms on the different sublattices.
These matrix elements are obtained as
\begin{eqnarray}
H_{AB}({\bf k})&=&t\Big(1+e^{i{\bf k}.\vec{\Delta}_{1}}+e^{i{\bf k}.\vec{\Delta}_{2}}\Big)+t^{"}
\Big(2cos(k_{y})+e^{-i\sqrt{3}k_{x}}\Big)\nonumber\\
&=&t\Big(1+2cos(k_{y}/2)e^{-i\sqrt{3}k_{x}/2}\Big)+t^{"}
\Big(2cos(k_{y})+e^{-i\sqrt{3}k_{x}}\Big),\nonumber\\
H_{AA}({\bf k})&=&2t'\Big(cos(\sqrt{3}k_{x}/2+k_{y}/2)+cos(\sqrt{3}k_{x}/2+k_{y}/2)+cos(k_{y}/2)\Big)\nonumber\\
&-&2\lambda \Big(sin\Big(\frac{1}{2}k_{y}\Big)-
sin\Big(\frac{\sqrt{3}}{2}k_{x}+\frac{1}{2}k_{y}\Big)-
sin\Big(\frac{\sqrt{3}}{2}k_{x}-\frac{1}{2}k_{y}\Big)\Big)-\mu-\sigma g\mu_{B}B,\nonumber\\
H_{BB}({\bf k})&=&-2t'\Big(cos(\sqrt{3}k_{x}/2+k_{y}/2)+cos(\sqrt{3}k_{x}/2+k_{y}/2)+cos(k_{y}/2)\Big)\nonumber\\
&+&2\lambda \Big(sin\Big(\frac{1}{2}k_{y}\Big)-
sin\Big(\frac{\sqrt{3}}{2}k_{x}+\frac{1}{2}k_{y}\Big)-
sin\Big(\frac{\sqrt{3}}{2}k_{x}-\frac{1}{2}k_{y}\Big)\Big)-\mu-\sigma g\mu_{B}B,\nonumber\\
H_{BA}({\bf k})&=&H^{*}_{AB}({\bf k}).
 \label{e9}
\end{eqnarray}
The matrix elements of ${\mathcal S}({\bf k})$, i.e. $S_{AA}({\bf k})$
, $S_{AB}({\bf k})$, $S_{BA}({\bf k})$
and $S_{BB}({\bf k})$ are expreseed as
\begin{eqnarray}
S_{AB}({\bf k})&=&s\Big(1+e^{i{\bf k}.\vec{\Delta}_{1}}+e^{i{\bf k}.\vec{\Delta}_{2}}\Big)+s^{"}
\Big(2cos(k_{y})+e^{-i\sqrt{3}k_{x}}\Big)\nonumber\\
&=&s\Big(1+2cos(k_{y}/2)e^{-i\sqrt{3}k_{x}/2}\Big)+s^{"}
\Big(2cos(k_{y})+e^{-i\sqrt{3}k_{x}}\Big)\nonumber\\
S_{AA}({\bf k})&=&1+2s'\Big(cos(\sqrt{3}k_{x}/2+k_{y}/2)+cos(\sqrt{3}k_{x}/2+k_{y}/2)+cos(k_{y}/2)\Big)\nonumber\\
S_{BB}({\bf k})&=&S_{AA}({\bf k})\;\;,\;\;S_{BA}({\bf k})=S^{*}_{AB}({\bf k}),
 \label{e10}
\end{eqnarray}
so that $s$ is the overlap between orbital wave function of electron respect to the nearest neighbor atoms,
$s'$ denotes the overlap between orbital wave function of electron respect to the next nearest neighbor atoms and
$s"$ implies the overlap between orbital wave function of electron respect to the next next nearest neighbor atoms.
The density functional theory and {\it ab initio} calculations has been determined the hopping amplitudes and overlap values $s,s',s"$
as\cite{xia}$t=-1.163,t'=-0.055,t"=-0.0836,s=0.01207,s'=0.0128,s"=0.048$.
These values for hopping amplitudes and overlap amounts have been obtained in a theoretical work\cite{chegel}.
Using the
Hamiltonian and overlap matrix forms
 in Eqs.(\ref{e9},\ref{e10}), the band structure of electrons, i.e. $E^{\sigma}_{\eta}({\bf k})$
has been found by solving equation
 $det\Big({\mathcal H}({\bf k})-E({\bf k}){\mathcal S}({\bf k})\Big)=0$.

 The final results for band structure
is lengthy and is not given here. The valence and condition bands of electrons have been presented by $E_{v}({\bf k})$ and $E_{c}({\bf k})$ respectively.
Using band energy spectrum,
the Hamiltonian in Eq.(\ref{e1}) can be rewritten by
\begin{eqnarray}
 H=\sum_{{\bf k},\sigma,\eta=c,v}
E^{\sigma}_{\eta}({\bf k})c^{\dag\sigma}_{\eta,{\bf k}}c^{\sigma}_{\eta,{\bf k}},
\label{e0.57}
\end{eqnarray}
where $c^{\sigma}_{\eta,{\bf k}}$ defines the creation operator of electron with spin
$\sigma$ in band index $\eta$ at wave vector ${\bf k}$.
The
electronic Green's function can be defined using the Hamiltonian in Eq.(\ref{e0.57}) as following expression
\begin{eqnarray}
G^{\sigma}_{\eta}({\bf k},\tau)=-\langle T_{\tau}
c^{\sigma}_{\eta,{\bf k}}(\tau)c^{\dag\sigma}_{\eta,{\bf k}}(0)\rangle,
 \label{e0.60}
\end{eqnarray}
where $\tau$ is imaginary time.
Using the model Hamiltonian in Eq.(\ref{e0.57}),
the Fourier transformations of Green's function is given by
\begin{eqnarray}
 G^{\sigma}_{\eta}({\bf k},i\omega_{n})=\int^{1/k_{B}T}_{0}d\tau e^{i\omega_{n}\tau}
G^{\sigma}_{\eta}({\bf k},\tau)=\frac{1}{i\omega_{n}-E^{\sigma}_{\eta}({\bf k})}.
\label{e0.61}
\end{eqnarray}
Here $\omega_{n}=(2n+1)\pi k_{B}T$ denotes the fermionic Matsubara frequency
in which $T$
is equilibrium temperature.
The electronic density of states of Germanene structure
 in the presence of intrinsic spin-orbit and external magnetic field
 can be obtained by electronic band structure as
\begin{eqnarray}
 Dos(E)=-\frac{1}{2N}Im\sum_{{\bf k},\sigma,\eta=v,c}\frac{1}{E-E^{\sigma}_{\eta}({\bf k})+i0^{+}}.
\label{e3.9}
\end{eqnarray}
Summation over wave vectors have been performed into first Brillouin zone of honeycomb lattice.
The density of states includes prominent asymmetric peaks due to the band edge of parabolic
subbands. The peaks positions arises from the band edge state energies and the density of states
heights are proportional to inverse square root of the sub band curvature and band degeneracy.
For determining the chemical potential, $\mu_{\sigma}$,
we use the relation between concentration of electrons ($n_{e}$) and chemical potential.
This relation is given by
\begin{eqnarray}
 n_{e}=\frac{1}{4N}\sum_{{\bf k},\eta,\sigma}\frac{1}{e^{E^{\sigma}_{\eta}({\bf k})/k_{B}T}+1}.
\label{e0.63}
\end{eqnarray}
Based on the values of
electronic concentration $n_{e}$, the chemical potential, $\mu$,
can be obtained by means Eq.(\ref{e0.63}).
\section{Optical absorption of Germanene structure}
The optical absorption of Germanene in the presence of spin-orbit coupling and
magnetic field are calculated.
In the following, the relation of the imaginary part of dielectric function, which is proportional to
the rate of photon absorption,
 is calculated for Germanene structure using Green's function method\cite{mahan}. When Germanene layer
 is excited by the electromagnetic field with polarization
of electric field along $x$ direction (see Fig.(1)) an Hamiltonian term as $\frac{e}{mc}{\bf A}.{\bf p}$ is
added to original Hamiltonian in Eq.(\ref{e1}).
 Kubo formulas allows
us the transverse dynamical conductivity ($\sigma_{xx}(\omega)$) in terms of correlation function between electrical currents
\cite{mahan} as
\begin{eqnarray}
\Re\sigma_{xx}(\omega)=
lim_{i\omega_{n}\longrightarrow\omega+i0^{+}}\frac{1}{\omega}\Im\Big[\int_{0}^{\beta}d\tau e^{i\omega_{n}\tau}
\langle T_{\tau}{\bf J}_{x}^{e}(\tau){\bf J}_{x}^{e}(0)\rangle\Big].
\label{e10.55}
\end{eqnarray}
The electrical current operator can be rewritten in terms of group velocity of electrons
 ${\bf v}_{\eta}({\bf k})=\nabla E^{\sigma}_{\eta}({\bf k})$ as follows
\begin{eqnarray}
{\bf J}_{x}^{e}=\sum_{{\bf k},\eta,\sigma}\frac{\partial E^{\sigma}_{\eta}({\bf k})}
{\partial k_{x}}c^{\dag\sigma}_{\eta,{\bf k}}c^{\sigma}_{\eta,{\bf k}}\cdot
 \label{e10.5}
\end{eqnarray}
The optical conductivity, as dynamical correlation function between
 electrical current operators, for single band model Hamiltonian
is obtained based on
Green's function using Wick's theorem.
The final results for
 the optical conductivity of Germanene monolayer
along $x$ zigzag direction are related to Green's function as
\begin{eqnarray}
\sigma(\omega)=\frac{e^{2}k_{B}}{4N}\sum_{{\bf k},\eta,\sigma}(\frac
{\partial E^{\sigma}_{\eta}({\bf k})}{\partial k_{x}})^{2}\int_{-\infty}^{\infty}
\frac{d\epsilon}{2\pi}\Big
(\frac{ n_{F}(\epsilon+\omega)-n_{F}(\epsilon)}{\omega}\Big)
\Big(2ImG^{\sigma}_{\eta}({\bf k},i\omega_{n}\longrightarrow\epsilon+i0^{+})\Big)^{2},
 \label{e5.66}
\end{eqnarray}
where $n_{F}(x)=\frac{1}{e^{x/k_{B}T}+1}$ is the Fermi-Dirac
distribution function and $T$ denotes the equilibrium temperature.
The imaginary part of dielectric function corresponding to the rate of
photon absorption by Germanene monolayer is obtained from
the dynamical conductivity via
\begin{eqnarray}
Im\epsilon(\omega)=\frac{4\pi}{\omega}Re\sigma(\omega).
 \label{3.4}
\end{eqnarray}
Substituting Eq.(\ref{e0.61}), i.e. electronic Green's function in band space, into Eq.(\ref{e5.6}) and
performing the numerical integration over wave vectors belonging to the
first Brillouin zone, the results of optical absorption are obtained.
Here, the contributions of both inter and intra band transitions on
the optical conductivity in Eq.(\ref{e5.6}) have been considered.
\section{Thermodynamic properties}
Transport properties such as electrical, thermal conductivities and thermoelectric coefficient
can be obtained by using band structure of electrons and spectral function in the system.
Using linear response theory, the transport coefficients under
the assumption of weak perturbing field, i.e weak gradient of temperature
and weak gradient of external electric potential are obtained.
The charge and thermal current are related to the
gradients $\nabla V$ and $\nabla T$ of the electric potential and the temperature
 by the following matrix relation
\begin{eqnarray}
\left(
                              \begin{array}{cccc}
                                {\bf J}_{1}\\
                                {\bf J}_{2} \\
  \end{array}
\right)=\left(
                              \begin{array}{cccc}
                                L_{11} & L_{12} \\
                                L_{21} & L_{22} \\
                                  \end{array}
\right)\left(
                              \begin{array}{cccc}
                                 {\bf E}\\
                                 -{\bf\nabla} T\\
                                  \end{array}
\right).
\label{e5.6.1.1}
\end{eqnarray}
${\bf J}_{1(2)}={\bf J}^{e}({\bf J}^{Q})$ implies
electrical (heat) current. Also $L_{ab}(a,b=1,2)$ are transport coefficients
which are determined by calculating
correlation function between the electrical and thermal current
operators. The off-diagonal thermoelectric coefficients are related to each other through the Onsager relation.
 Based on spectral function, i.e. the imaginary part of retarded Green's function,
one can calculate the static transport coefficients
 of single layer Germanene along $x$ direction as
\begin{eqnarray}
L_{11}&=&\frac{e^{2}k_{B}T}{4N}\sum_{{\bf k},\eta,\sigma}(\frac
{\partial E^{\sigma}_{\eta}({\bf k})}{\partial k_{x}})^{2}\int_{-\infty}^{\infty}\frac{d\epsilon}{2\pi}\Big
(\frac{-\partial n_{F}(\epsilon)}{\partial\epsilon}\Big)
\Big({\mathcal A}_{\eta}^{\sigma}({\bf k},\epsilon)\Big)^{2},\nonumber\\
L_{12}&=&\frac{ek_{B}T}{4N}\sum_{{\bf k},\eta,\sigma}(\frac
{\partial E^{\sigma}_{\eta}({\bf k})}{\partial k_{x}})^{2}\int_{-\infty}^{\infty}\frac{d\epsilon}{2\pi}\epsilon\Big
(\frac{-\partial n_{F}(\epsilon)}{\partial\epsilon}\Big)
\Big({\mathcal A}_{\eta}^{\sigma}({\bf k},\epsilon)\Big)^{2},\nonumber\\
L_{22}&=&\frac{ek_{B}T}{4N}\sum_{{\bf k},\eta,\sigma}(\frac
{\partial E^{\sigma}_{\eta}({\bf k})}{\partial k_{x}})^{2}\int_{-\infty}^{\infty}\frac{d\epsilon}{2\pi}\epsilon^{2}\Big
(\frac{-\partial n_{F}(\epsilon)}{\partial\epsilon}\Big)
\Big({\mathcal A}_{\eta}^{\sigma}({\bf k},\epsilon)\Big)^{2},
 \label{e5.6}
\end{eqnarray}
where $n_{F}(x)$ the Fermi-Dirac
distribution function. Moreover ${\mathcal A}_{\eta}^{\sigma}({\bf k},\epsilon)\equiv-2ImG^{\sigma}_{\eta}
({\bf k},i\omega_{n}\longrightarrow\epsilon+i0^{+})$ denotes the electric spectral function.
Substituting electronic Green's function into Eq.(\ref{e5.6}) and
performing the numerical integration over wave vector through
first Brillouin zone, the results of transport
coefficients have been obtained.
 In the presence of a temperature gradient ($\nabla T$)
and in open circuit situation, i.e.${\bf J}^{e}=0$,
 heat current is related
to temperature gradient via
${\bf J}^{Q}=\kappa\nabla T$ where $\kappa$ is the thermal conductivity and is obtained using
transport coefficients as\cite{grosso}
 \begin{eqnarray}
 \kappa&=&\frac{1}{T^2}(L_{22}-\frac{L_{12}^{2}}{L_{11}}).
 \label{e3.8}
 \end{eqnarray}
The ratio of the measured voltage to the temperature gradient applied across the sample is known
as the Seebeck coefficient (or the thermopower) and is given by $S=\nabla V/\nabla T$, where
$\nabla V$ is the potential difference between two points of the sample\cite{grosso}. In linear response
approximation the Seebeck coefficient is related to transport coefficients as
\begin{eqnarray}
S=-\frac{1}{T}\frac{L_{12}}{L_{11}}.
\label{e3.99}
\end{eqnarray}
 $S$ denotes the thermopower which describe the voltage generation owing to the temperature gradient.
The sign of $S$ determines the sign of majority carriers of thermal transport in the Germanene structure.
 Moreover the static
 conductivity can be obtained by taking the zero limit frequency of dynamical electrical conductivity
, i.e.$\sigma(T)=\lim_{\omega\longrightarrow0}\sigma(\omega)$. After
some algebraic calculations, the final result for the static electrical conductivity of
 monolayer Germanene in the presence of spin-orbit coupling and magnetic field
along $x$ direction is related to the spectral function as
\begin{eqnarray}
\sigma(T)=\frac{e^{2}k_{B}}{4N}\sum_{{\bf k},\eta,\sigma}(\frac
{\partial E^{\sigma}_{\eta}({\bf k})}{\partial k_{x}})^{2}\int_{-\infty}^{\infty}\frac{d\epsilon}{2\pi}\Big
(\frac{-\partial n_{F}(\epsilon)}{\partial\epsilon}\Big)
\Big({\mathcal A}_{\eta}^{\sigma}({\bf k},\epsilon)\Big)^{2},\nonumber\\
 \label{e5.666}
\end{eqnarray}

 The electronic specific heat could be obtained by means of Green's function as
\begin{eqnarray}
 C(T)&=&\int^{\infty}_{-\infty}dE ED(E)\frac{\partial f(E,T)}{\partial T}
\nonumber\\&=&-\frac{k_{B}}{2\pi N(k_{B}T)^{2}}\sum_{{\bf k},\sigma}\sum_{\eta=v,c}\int_{-\infty}^{+\infty}dE
\frac{E^{2}e^{E/(k_{B}T)}}{(1+e^{E/(k_{B}T)})^{2}}Im G^{\sigma}_{\eta}({\bf k},E).
\label{e2589}
\end{eqnarray}
In the next section, the numerical results of electronic properties of single layer Germanene are presented.
\section{Numerical Results and Discussions}
Here we present our numerical results for the electronic properties of Germanene layer in the presence
of magnetic field and spin-orbit coupling effects. Using band structure of electron, we can obtain the electronic Green's function
in Eq.(\ref{e0.61}). Afterwards the optical absorption, transport coefficients and specific heat are found by substitution of
Green's function into Eqs.(\ref{e5.66},\ref{e5.6},\ref{e2589}), respectively.
Finally static electrical conductivity, thermal conductivity and thermopower are readily calculated based on
 Eqs.(\ref{e3.8},\ref{e3.99},\ref{e5.666}). Also the energy dependence of density of states is obtained using Eq.(\ref{e3.9}).
 Both inter and intra band
transitions contribute to results of electronic properties of Germanene monolayer.
 Also the calculation is performed within full
Brillouin zone beyond Dirac cone approximation.

The optimized atomic structure of the Germanene with primitive unit cell vector length $a=1$
is shown in Fig.(1). The primitive unit cell include two Ge atoms.
The total density of states ($Dos(E)$)
of single layer Germanene in
the absence of magnetic field for different values of
spin-orbit coupling strength $\lambda/t=0.1,0.2,0.3,0.4,0.5$ has been shown in Fig.(3). The effects of next nearest
neighbor and next next nearest neighbor hopping integrals shows that density of states curves shows no symmetry around $E/t\approx0.2$
 according
to Fig.(3). This arises from this fact that model Hamiltonian loses the particle-hole symmetry
in the presence of next and next next nearest neighbor hopping integrals, $t',t"\neq0.0$.
DOS curves obtained from Kane-Mele model due to spin-orbit coupling have two Van Hove singularities in the conduction and valence
area on the both sides of the Fermi level and in energies correspond to the lowest states in the band structure.
The DOS indicates a linear behavior in terms of energy around the minimum point $E/t\approx-0.2$ and reaches to peaks in both sides of
Fermi level at $E=0.0$. Such behavior arises from the linear dispersion properties of band structure around Dirac points of
the first Brillouin zone.
The curves of density of states includes sharp peaks at energy values $E/t=-1.0,1.0,3.0$.
For undoped case Fermi energy has zero value.
In addition two main peaks in DOS at $E/t\approx-1.0,1.0$, there are other peaks in $E/t=-3.0,3.0$ for $\lambda/t=0.5$.
In other words, these additional peaks become more clear upon increasing spin-orbit coupling strength.
Another novel feature in Fig.(3) is the decrease of DOS at Fermi energy with enhancement of $\lambda$ as shown in Fig.(3).
In fact the spin-orbit coupling leads to increase insulating property of Germanene layer.
However our numerical results for density of states of Germanene layer in the absence of spin-orbit coupling is in agreement
with DFT results, there is no study on density of states of Germanene layer in the presence of spin-orbit coupling in the context of DFT method.
The linear behavior of density of states of Germanene at low value of spin-orbit coupling in Fig.(3) of the manuscript is in good agreement
with density of states results of Germanene by using DFT method\cite{chegel}.

We have studied the effect of magnetic field on density of states of single layer Germanene
 structure due to spin-orbit coupling.
In Fig.(4) we have plotted D(E)
versus normalized energy $E/t$ for different magnetic field values, namely
$g\mu_{B}B/t=0.0,0.2,0.4,0.6$ at fixed value for spin-orbit coupling $\lambda/t=0.4$.
Two novel features are pronounced in this figure. Upon turning the magnetic field $B$, double van hove singularities
in density of states is observed for each positive or negative energy region.
The energy difference between two sharp peaks in density of states
in positive or negative energy region increases with magnetic field due to Zeeman splitting.
In other hand the metallic property of Germanene plane enhances with magnetic field according to Fig.(4).
This arises from this fact that density of states value in chemical potential value increases with magnetic field and consequently
the conductivity of Germanene layer enhances. In undoped case, the chemical potential for undoped graphene is located at zero energy.

We begin
the investigations of thermal properties of Germanene
layer with a discussion of the thermal conductivity in the presence
of magnetic field and spin-orbit coupling effects.
In Fig.(5) results for the thermal conductivity ($\kappa(T)$) of undoped Germanene structure
are presented versus normalized temperature ($kT/J$, where $k$ is the
Boltzmann constant) for different spin-orbit coupling values in the absence of magnetic field.
 Several features are remarkable. Each curve
shows an increasing behavior at low temperatures which manifests the presence
of a finite-energy gap in the energy spectrum.
The variation of $\lambda$ has no considerable effect on temperature behavior of thermal conductivity according to Fig.(5).
 There
is also a finite temperature maximum in the
thermal conductivity so that its temperature position moves to higher temperature upon
increasing $\lambda/t$. In addition, at fixed value of temperature above peak position, thermal conductivity of Germanene layer increases
with spin-orbit coupling constant.
 Below the characteristic
temperature of the maximum, the enhancement of temperature leads to
increase the rate of transition of electrons to the excited state. Therefore we see an
increase of the thermal
conductivity at low temperatures, see Fig.(5). Upon further increase
of the temperature, the electrons suffer from scattering effects on each other
which reduces the thermal conductivity. Hence the temperature dependence of each
curve is due to competition between the two phenomena, the increase of the
transition rate of electrons from ground state to excited one within the physical limits and the scattering of the electrons
at higher temperatures.

We also studied the effect of the magnetic field values, i.e. $g\mu_{B}B/t$,
on the temperature dependence of the thermal conductivity of the Germanene layer.
 In Fig.(6) we plot $\kappa$ versus normalized
temperature for several values of the longitudinal magnetic field, namely
$g\mu_{B}/t=0.0,0.2,0.4,0.6,0.8$ under the half filling constraint for $\lambda/t=0.0$.
The half filling case means the average of number of electrons per atomic site is unit. In other words the electronic concentration is unit.
This plot shows the increase of
the thermal conductivities with decrease of the anisotropy parameter in temperature region $0.5<k_{B}T/t<1.5$.
However thermal conductivity rises with magnetic field at fixed temperature in the regions $k_{B}T/t>1.5$ and $0<k_{B}T/t<0.5$.
The height of peak in thermal conductivity reduces with magnetic field
although the temperature position of peak moves to higher amounts as shown
in Fig.(6).

In Fig.(7) we present the magnetic field dependence of the thermal
conductivity of doped Germanene for various chemical potential $\mu/t$ for fixed $kT/t=0.04$ and
$\lambda/t=0.0$.
Upon increasing chemical potential, the peak in temperature dependence of thermal
conductivity becomes more obvious.
However thermal conductivity for chemical potential values $\mu/t=0.3,0.4$
does not include any peak.
Magnetic field dependence of thermal conductivity presents a decreasing behavior
 at magnetic fields above
 2.5 for all values of chemical potentials.
In addition, at fixed value of magnetic field,
thermal conductivity enhances with electron doping.
 This increase of the conductivity with chemical potential
as seen in Fig.(7) can be
understood from the enhancement of transition rate of electrons form valence band to conduction
one.

We have studied the static electrical conductivity of Germanene layer due to magnetic field
and spin-orbit coupling strength.
The resulting of the electrical conductivity of undoped Germanene layer
as a function of normalized temperature $k_{B}T/t$ for different values of
spin-orbit coupling constant at $g\mu_{B}B/t=0$ has been plotted in Fig.(8).
 This figure indicates the
increasing behavior for temperature dependence of
 electrical conductivity is clearly observed in temperature region
$k_{B}T/t<1.0$ for all amounts of $\lambda/t$.
This peak arises from band gap in density of states of Germanene due to spin-orbit coupling.
In addition, at fixed value of normalized temperature, lower spin-orbit coupling causes less
 band gap in density of states and consequently higher values in electrical conductivity.
Another novel feature is the independency of electrical conductivity on $\lambda$
at low normalized temperatures below 0.4 according to Fig.(8)

In Fig.(9) we present the electrical conductivity of the Germanene plane versus
normalized temperature, $k_{B}T/t$ for different values of magnetic field, namely
$g\mu_{B}B/t=0.2,0.3,0.4,0.5,0.6,0.7$.
 in half filling case at fixed $\lambda/t=0.4$.
 According to Fig.(4), the density of states at Fermi energy increases with
magnetic field and consequently magnetic field improves
the conductance property of the Germanene layer.
Therefore the conductivity increases with enhancement of magnetic field
at fixed low temperatures. For each value of magnetic field, the electrical conductivity
increases with temperature at low temperatures which comes form the presence
of a finite energy gap in the density of states.
Moreover lower vales of magnetic field present more rapid increase at low
 temperatures corresponding
to smaller energy gap as shown in Fig.(9).
There is also a peak in the electrical conductivity which moves to lower
temperature upon increasing magnetic field. The temperature behavior of electrical conductivity
at higher values show the conductivity is independent of magnetic field and all plots
fall on each other on the whole range of temperature above normalized value 1.0.

In the absence of spin-orbit coupling,
we have plotted the electrical conductivity of doped Germanene layer as a function of normalized magnetic field for
different chemical potentials, namely $\mu/t=0.0,1.0,1.1,1.2$ has been shown
in Fig.(10). The temperature has been fixed at $k_{B}T/t=0.04$.
There is a peak in conductivity around normalized magnetic field 1.5 for all chemical
potential values. In addition, at fixed magnetic field, the electrical conductivity
enhances upon increasing chemical potential. This fact can be understood from this point
the increase of chemical potential leads to raise the population of electrons which
increases the electrical conductivity.

The effects of spin-orbit coupling strengths on magnetic field dependence of electrical conductivity of Germanene layer at $k_{B}T/t=0.04$ in the
half filling case have been plotted in Fig.(11).
This figure shows the electrical conductivity rises with magnetic field at magnetic field values below 2.0.
Upon more increasing magnetic field above 2.0, electrical conductivity decreases with increase of magnetic field. The peak in conductivity
moves to higher magnetic fields amounts with enhancement of $\lambda/t$. Moreover, at fixed magnetic field, the energy gap decreases with
spin-orbit coupling strength and consequently the conductivity reduces. Also the effects of spin-orbit coupling is more clear at higher magnetic
field values rather than lower values.

Considering magnetothermal effects using Eq.(\ref{e5.6.1.1}),
the Seebeck coefficient $S$ under the
condition of zero electrical current ${\bf J}^{e}=0$ and for ballistic transport Eq.(\ref{e3.9}).
In Fig.(12) we depict the
Seebeck coefficient $S$ of doped monolayer Germanene
as a function of normalized temperature $k_{B}T/t$
for several values of normalized spin-orbit coupling $\lambda/t$ at zero magnetic field with $\mu/t=0.2$.
We note that absolute value of Seebeck coefficient of Germanene layer for non zero spin orbit coupling increases with $\lambda$
at fixed temperature. The sign of $S$ for non zero spin orbit coupling is negative however the Seebeck coefficient gets positive sign in the
absence of spin-orbit coupling.
In Ref.(\cite{furu}), it was suggested that the sign of
$S$ is a criterion to clarify the types of carriers; a positive (negative) $S$ implies that the
charge and heat are dominantly carried by electrons (holes).
Fig.(12) indicates the Seebeck coefficient tends to zero for all $\lambda/t$ at temperatures above normalized value 1.0.
The peak in Seebeck coefficient appears at $k_{B}T/t\approx0.125$ for all values of $\lambda$ according to Fig.(12).
The height of peak in $S$ for non zero values of $\lambda$ enhances with spin-orbit coupling strength.

The dependence of Seebeck coefficient of undoped monolayer graphene on the temperature
 for different
magnetic fields at zero spin-orbit coupling has been plotted
in Fig.(13).
The sign of Seebeck coefficient is negative for $g\mu_{B}B/t=0.4,0.6$ on the whole range of temperature and its sign
becomes positive for $g\mu_{B}B/t=0.7,0.8$.
However there is a peak in temperature dependence of
 the absolute value of Seebeck coefficient for all magnetic fields around $k_{B}T/t=0.25$.
The Seebeck coefficient curves fall on each other at higher normalized temperatures above 1.5 for all magnetic fields.

Fig.(14) shows specific heat $C_{v}$ of Germanene as a function of normalized temperature $k_{B}T/t$ for different $\lambda$
in the half filling case at zero magnetic field. The specific heat is zero at the zero temperature and it increases with temperature
up to maximum point of the curve.
 Larger values of $\lambda/t$ show lower value for specific heat at low temperatures corresponding
to larger energy gap.
Indeed the low temperature limit of specific heat is proportional to
$1/Te^{-E_{g}/T}$ and therefore the decrease of specific heat with energy gap can be understood.
 The peak in specific heat goes to higher temperature with increasing spin-orbit coupling constant
according to Fig.(14).
In temperature region above normalized value 1.5, the increase of $\lambda$ leads to enhance the
specific heat.

We have examined the effects of magnetic field on temperature dependence of
specific heat of undoped Germanene layer in the absence of spin orbit interaction and
the numerical results have been plotted in Fig.(15).
All curves have a peak so that temperature position of peak moves to higher temperature
due to enhancement of magnetic field. The specific heat is not remarkably affected by magnetic field
for temperatures above 1.5 as shown in Fig.(15).

One of the important aim in this work is investigation of
the optical properties of Germanene and for this purpose we should use the relation
Eqs.(\ref{e5.66},\ref{3.4}).
Fig.(16) shows the imaginary part of dielectric function, i.e. the absorption rate , of
Germanene in the presence of spin orbit interaction in the half filling case at constant
temperature $k_{B}T/t=0.05$ at zero magnetic field.
The optical absorption of Germanene layer for $\lambda/t=0.2,0.25,0.35$ displays a
Drude response at low frequency due to intraband transitions and a flat interband absorption
which commences at photon frequency $\omega/t\approx1.25$. Such general behavior has been
predicted for Graphene monolayer and bilayer graphene in the other theoretical work\cite{tarbetArxiv}.
The Drude response disappears at higher spin orbit coupling $\lambda/t=0.6,0.7$
 where the sample develops insulating behavior due to high values of spin orbit interaction
strength.
The optical absorption rate decreases with photon frequency above peak position.

Now the influence of chemical potential on photon frequency dependence of
 imaginary part of dielectric function of doped Germanene
layer at $\lambda/t=0.5$ for $k_{B}T/t=0.05$ is ivestigated in Fig.(17).
The Drude weight becomes more considerable and gets the higher values with
increasing the chemical potential
$\mu/t$. This fact can be understood from this point that the metallic property of Germanene
has been improved due to electron doping and consequently the Drude weight increases.
The flat peak in optical absorption appears at finite frequency $\omega/t=1.25$.
The height of peak increases with chemical potential since the transition rate of electrons from
valence band to conduction one enhances with electron concentration according to Fig.(17).
Moreover the optical absorption of Germanene layer is less chemical potential dependent at
frequencies above 2.0 as shown in Fig.(17).

Finally we have studied the frequency dependence of $Im\epsilon(\omega)$ of
Germanene layer for different values of magnetic field $g\mu_{B}B/t$ at $\lambda/t=0.5$
for $k_{B}T/t=0.05$ in Fig.(18).
According to energy dependence of $Dos(E)$ in Fig.(4), the metallic property of
Germanene layer increases with magnetic field and consequently the Drude response of the
Germanene layer enhances with $g\mu_{B}B/t$.
Moreover the frequency position of flat peak in optical absorption moves to lower amounts
with increasing magnetic field however the height of peak position is not considerably affected
by magnetic field according to Fig.(18).
Another novel feature in Fig.(18) is clearly observed.
The frequency dependence of optical absorption is independent of magnetic field in the frequency
region $\omega/t>1.25$ and all curves fall on each other on the range of frequency
 $\omega/t>1.25$.

It is remarkable to add few comments concerning the zero frequency limit of imaginary part of dielectric function in Figs.(16,17,18).
According to our numerical results for optical absorption in Figs.(16,17,18), it is clearly observed that there is Drude weight for the
imaginary part of the dielectric function. This Drude weight is the zero frequency limit of imaginary part of the dielectric function
where we find the non zero value for $Im\epsilon$. This non zero value is an evidence for metallic property of the structure in suitable parameter values.
Although we find the zero value for imaginary part of the dielectric function at zero frequency limit for $\lambda/t=0.6,0.7$ where the structure behaves as
semiconductor.

\begin{figure}
\begin{center}
\epsfxsize=0.8\textwidth
\includegraphics[width=12.cm]{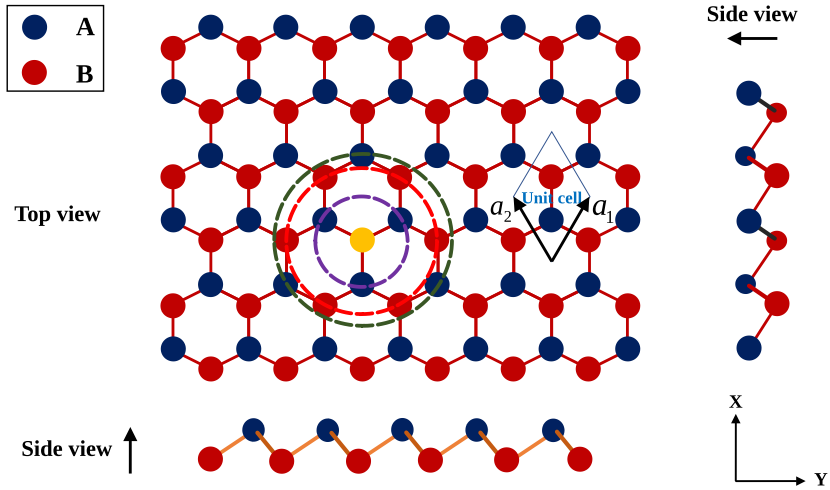}
\small
\begin{flushleft}
\caption{Crystal structure of Germanene}
\end{flushleft}
\end{center}
\end{figure}
\begin{figure}
\begin{center}
\epsfxsize=0.8\textwidth
\includegraphics[width=12.cm]{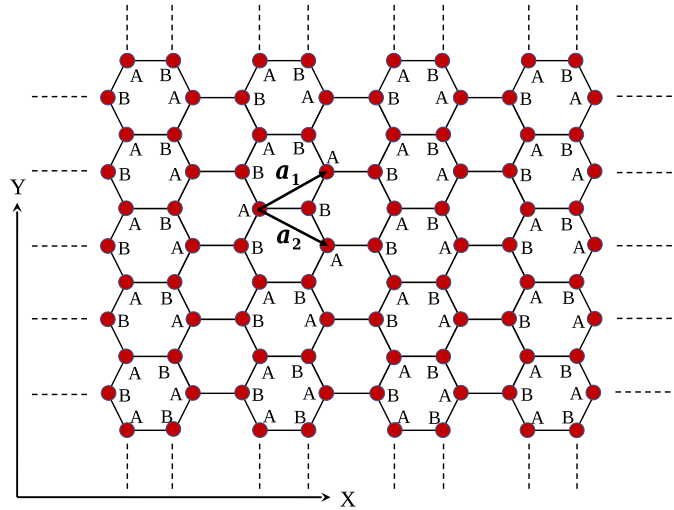}
\small
\begin{flushleft}
\caption{The structure of honeycomb structure is shown. The
light dashed lines denote the Bravais lattice unit cell. Each cell
includes two nonequivalent sites, which are indicated by A and B.
${\bf a}_{1}$ and ${\bf a}_{2}$ are the primitive vectors of unit
cell.}
\end{flushleft}
\end{center}
\end{figure}
\begin{figure}
\begin{center}
\epsfxsize=0.8\textwidth
\includegraphics[width=12.cm]{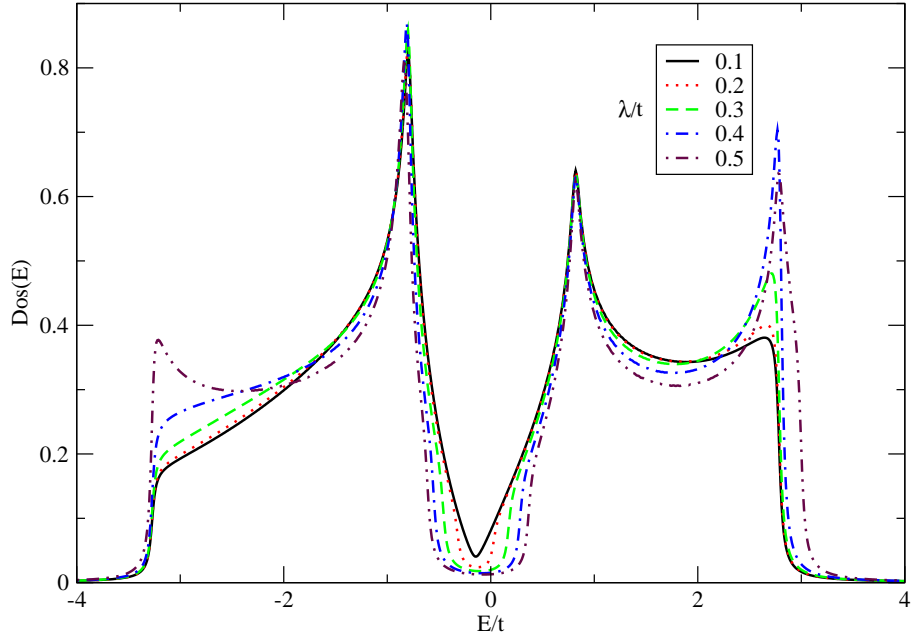}
\small
\begin{flushleft}
\caption{Density of states (DOS(E)) of Germanene as a function of energy for different values of
normalized spin-orbit coupling strength $\lambda/t$ in the absence of magnetic field for half filling case }
\end{flushleft}
\end{center}
\end{figure}
\begin{figure}
\begin{center}
\epsfxsize=0.8\textwidth
\includegraphics[width=12.cm]{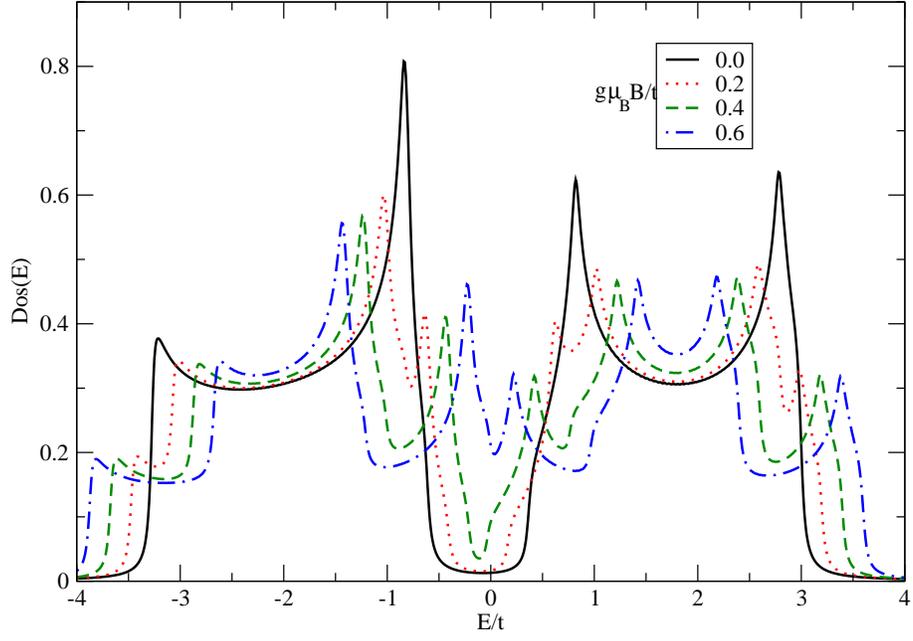}
\small
\begin{flushleft}
\caption{Density of states (DOS(E)) of Germanene as a function of energy for different values of
normalized magnetic field $g\mu_{B}B/t$ at $\lambda/t=0.4$ for half filling case.}
\end{flushleft}
\end{center}
\end{figure}
\begin{figure}
\begin{center}
\epsfxsize=0.8\textwidth
\includegraphics[width=12.cm]{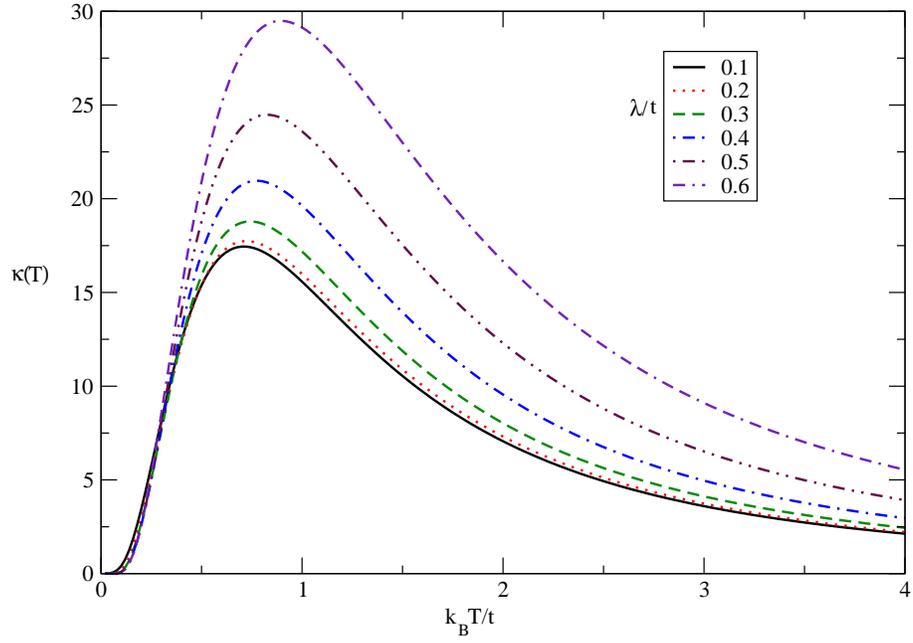}
\small
\begin{flushleft}
\caption{Thermal conductivity ($\kappa(T)$) of Germanene as a function of normalized temperature $k_{B}T/t$ for different values of
normalized spin-orbit coupling strength $\lambda/t$ in the absence of magnetic for half filling case.}
\end{flushleft}
\end{center}
\end{figure}
\begin{figure}
\begin{center}
\epsfxsize=0.8\textwidth
\includegraphics[width=12.cm]{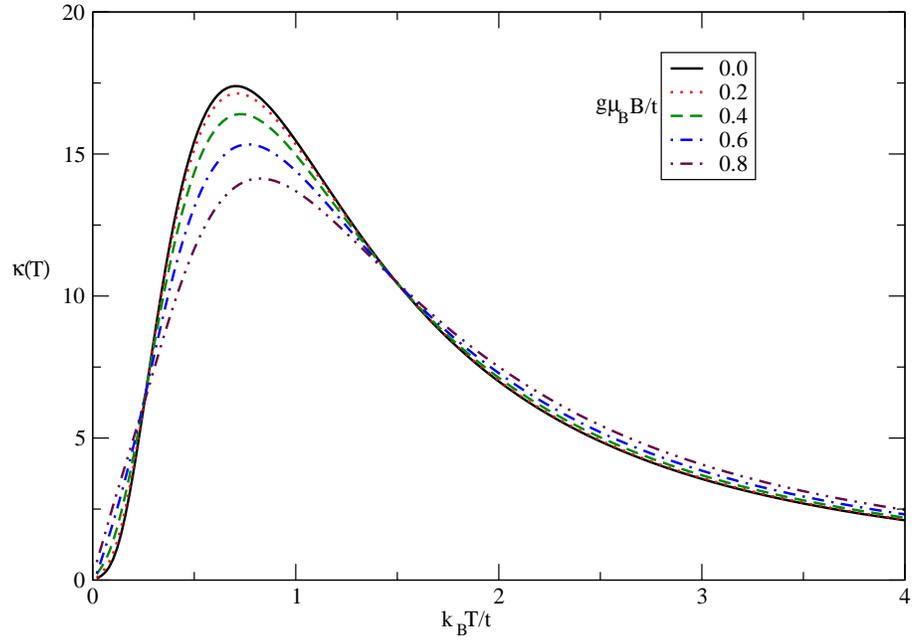}
\small
\begin{flushleft}
\caption{Thermal conductivity ($\kappa(T)$) of Germanene as a function of normalized temperature $k_{B}T/t$ for different values of
normalized magnetic field $g\mu_{B}B/t$ at $\lambda/t=0.0$ for half filling case.}
\end{flushleft}
\end{center}
\end{figure}
\begin{figure}
\begin{center}
\epsfxsize=0.8\textwidth
\includegraphics[width=12.cm]{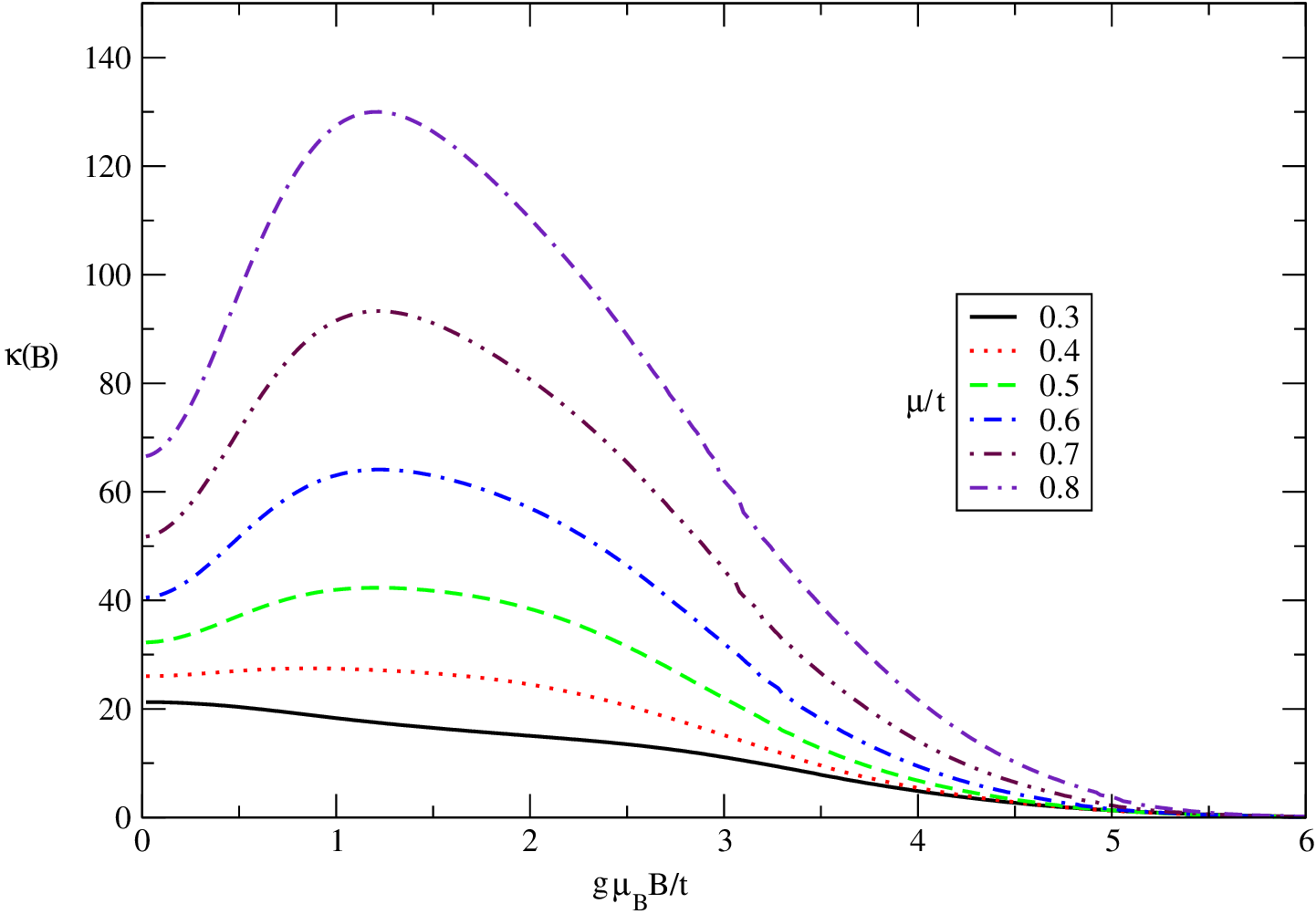}
\small
\begin{flushleft}
\caption{Thermal conductivity ($\kappa(B)$) of Germanene as a function of normalized magnetic field $g\mu_{B}B/t$
 for different values of normalized chemical potential $\mu/t$
 at $\lambda/t=0.0$ for $k_{B}T/t=0.04$.}
\end{flushleft}
\end{center}
\end{figure}
\begin{figure}
\begin{center}
\epsfxsize=0.8\textwidth
\includegraphics[width=12.cm]{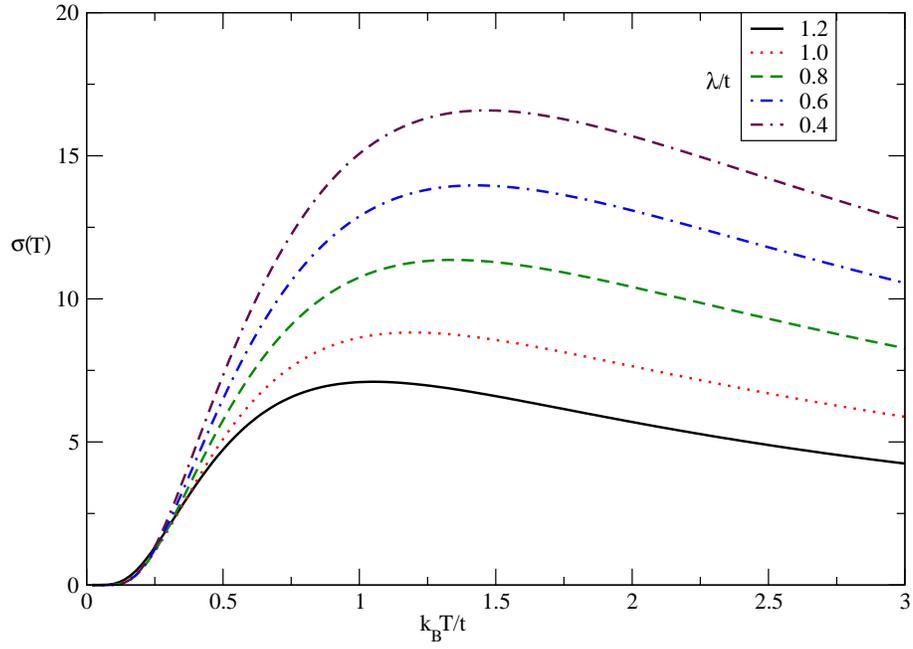}
\small
\begin{flushleft}
\caption{Electrical conductivity ($\sigma(T)$) of Germanene as a function of normalized temperature $k_{B}T/t$ for different values of
normalized spin-orbit coupling strength $\lambda/t$ in the absence of magnetic for half filling case.}
\end{flushleft}
\end{center}
\end{figure}
\begin{figure}
\begin{center}
\epsfxsize=0.8\textwidth
\includegraphics[width=12.cm]{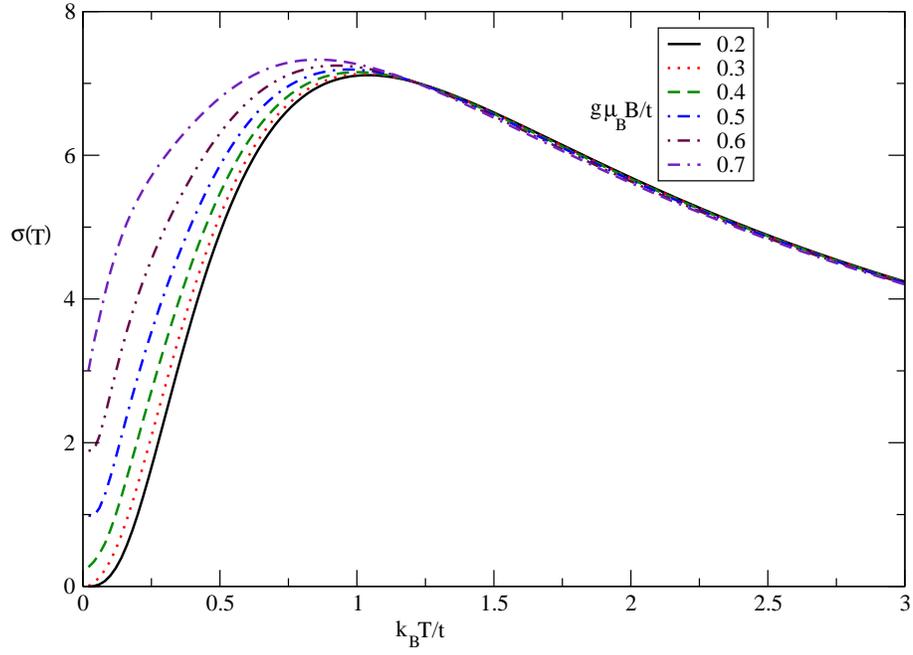}
\small
\begin{flushleft}
\caption{Electrical conductivity ($\sigma(T)$) of Germanene as a function of normalized temperature $k_{B}T/t$ for different values of
normalized magnetic field $g\mu_{B}B/t$ at $\lambda/t=0.4$ for half filling case.}
\end{flushleft}
\end{center}
\end{figure}
\begin{figure}
\begin{center}
\epsfxsize=0.8\textwidth
\includegraphics[width=12.cm]{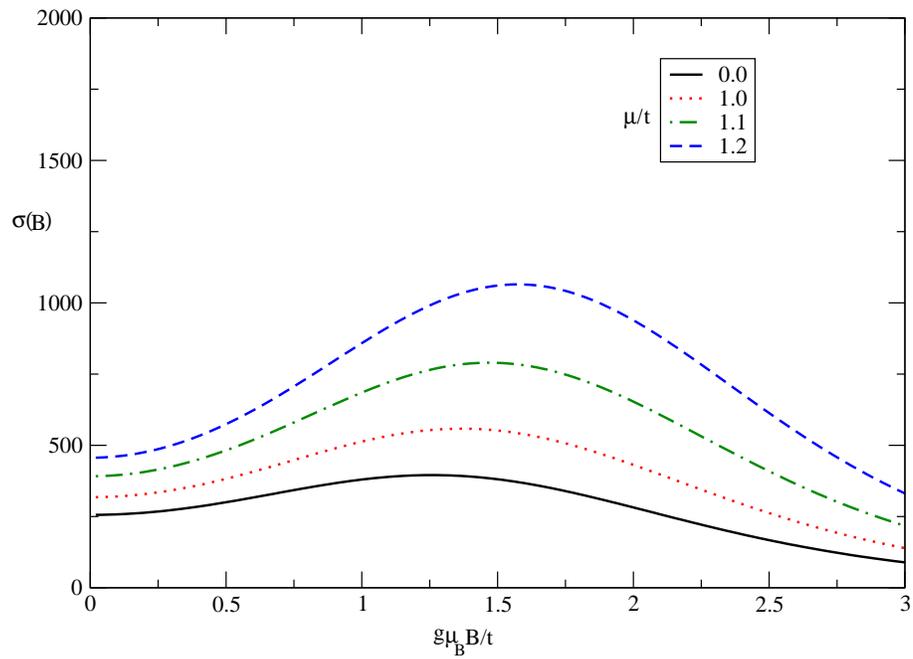}
\small
\begin{flushleft}
\caption{Electrical conductivity ($\sigma(B)$) of Germanene as a function of normalized magnetic field $g\mu_{B}B/t$
 for different values of normalized chemical potential $\mu/t$
 at $\lambda/t=0.0$ for $k_{B}T/t=0.04$.}
\end{flushleft}
\end{center}
\end{figure}
\begin{figure}
\begin{center}
\epsfxsize=0.8\textwidth
\includegraphics[width=12.cm]{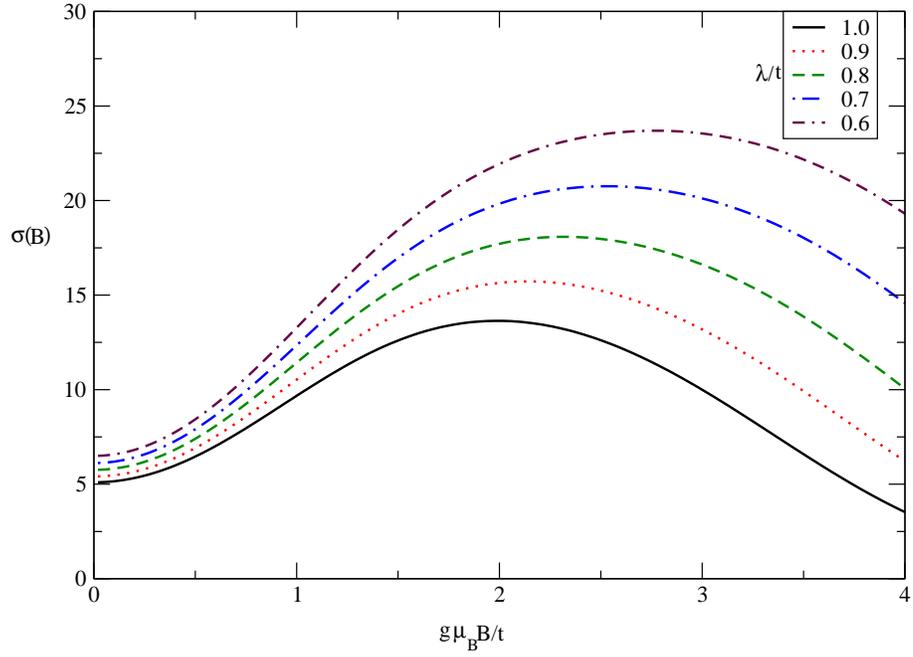}
\small
\begin{flushleft}
\caption{Electrical conductivity ($\sigma(B)$) of Germanene as a function of normalized magnetic field $g\mu_{B}B/t$
 for different values of normalized spin-orbit coupling $\lambda/t$
 at $\mu/t=0.0$ for $k_{B}T/t=0.04$.}
\end{flushleft}
\end{center}
\end{figure}
\begin{figure}
\begin{center}
\epsfxsize=0.8\textwidth
\includegraphics[width=12.cm]{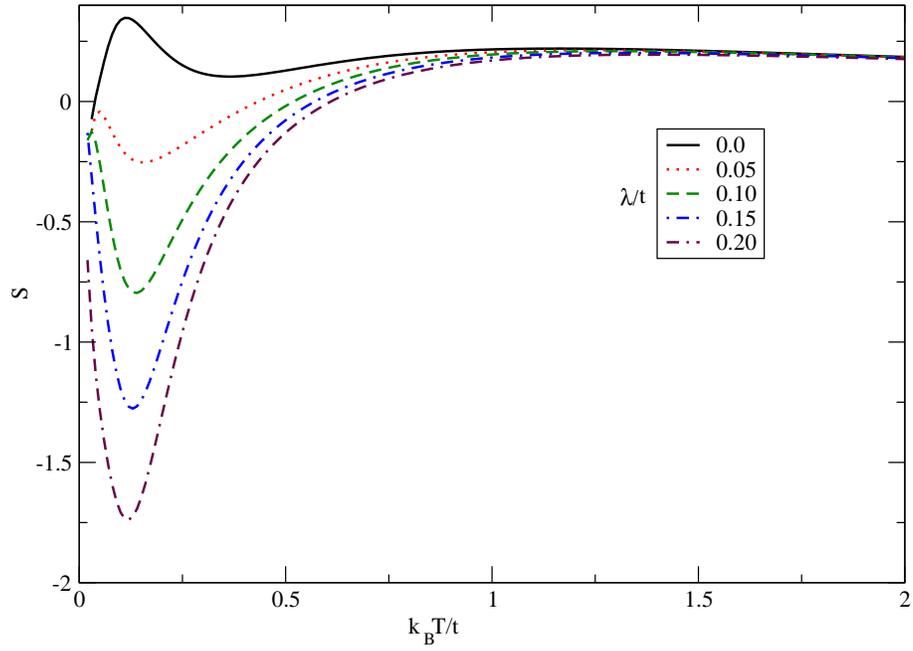}
\small
\begin{flushleft}
\caption{Seebeck coefficient ($S$) of Germanene as a function of normalized temperature $k_{B}T/t$
 for different values of normalized spin-orbit coupling $\lambda/t$
 at $\mu/t=0.2$ in the absence of magnetic field.}
\end{flushleft}
\end{center}
\end{figure}
\begin{figure}
\begin{center}
\epsfxsize=0.8\textwidth
\includegraphics[width=12.cm]{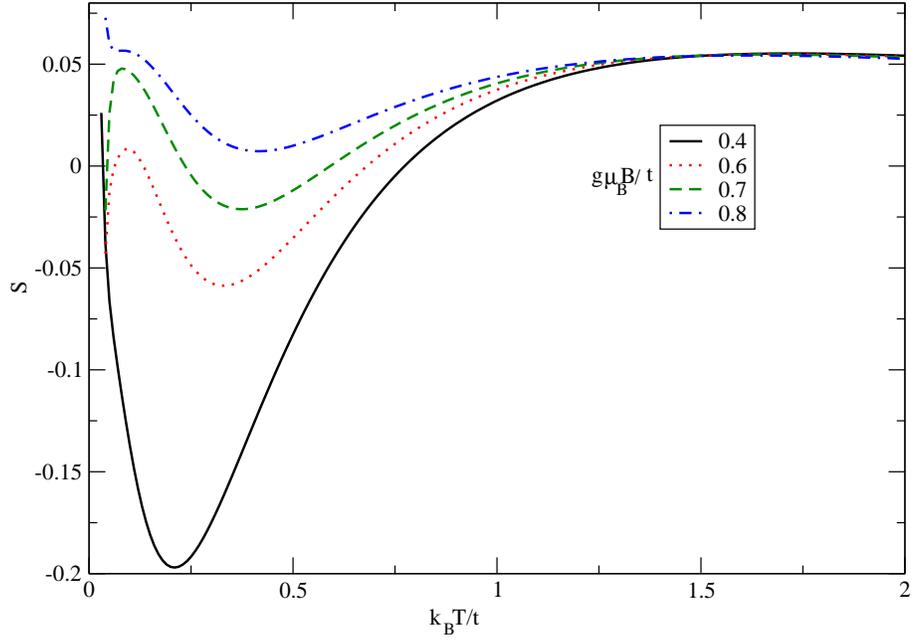}
\small
\begin{flushleft}
\caption{Seebeck coefficient ($S$) of Germanene as a function of normalized temperature $k_{B}T/t$
 for different values of normalized magnetic field $g\mu_{B}B/t$
 at $\mu/t=0.0$ in the absence of spin-orbit coupling constant.}
\end{flushleft}
\end{center}
\end{figure}

\begin{figure}
\begin{center}
\epsfxsize=0.8\textwidth
\includegraphics[width=12.cm]{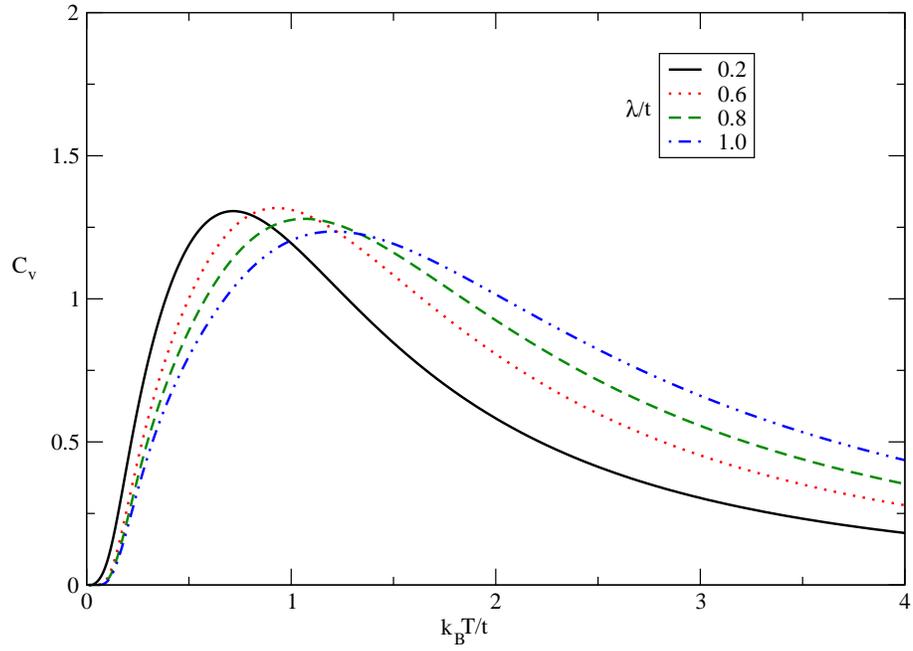}
\small
\begin{flushleft}
\caption{Specific heat ($C_{v}$) of Germanene as a function of normalized temperature $k_{B}T/t$ for different values of
normalized spin-orbit coupling strength $\lambda/t$ in the absence of magnetic for half filling case.}
\end{flushleft}
\end{center}
\end{figure}
\begin{figure}
\begin{center}
\epsfxsize=0.8\textwidth
\includegraphics[width=12.cm]{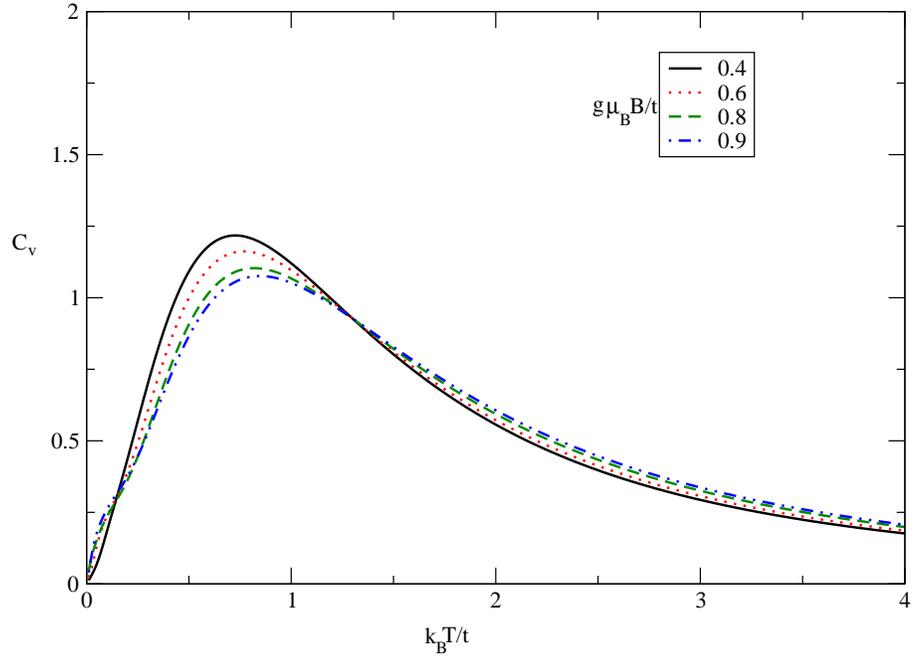}
\small
\begin{flushleft}
\caption{Specific heat ($C_{v}$) of Germanene as a function of normalized temperature $k_{B}T/t$ for different values of
normalized magnetic field $g\mu_{B}B/t$ at $\lambda/t=0.0$ for half filling case.}
\end{flushleft}
\end{center}
\end{figure}
\begin{figure}
\begin{center}
\epsfxsize=0.8\textwidth
\includegraphics[width=12.cm]{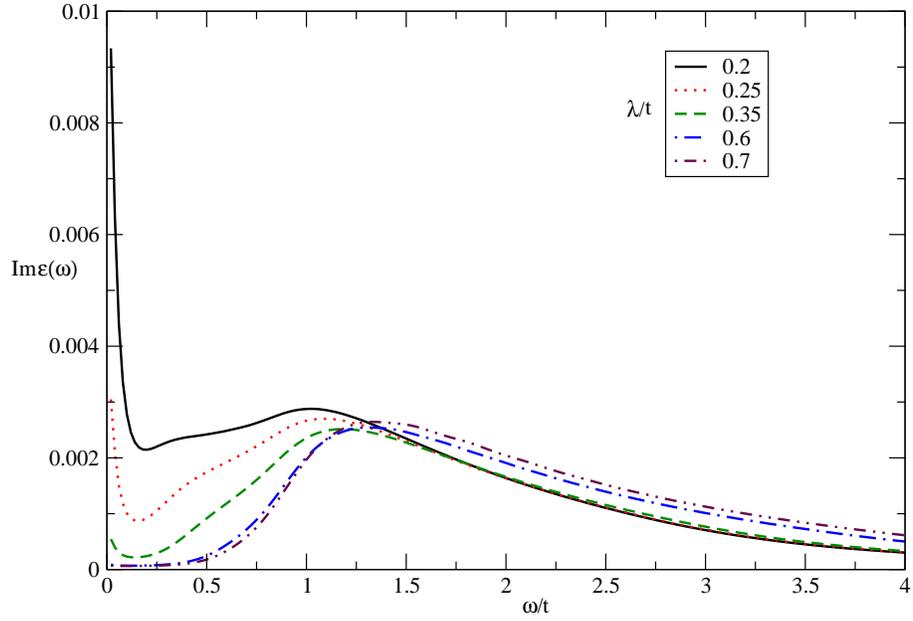}
\small
\begin{flushleft}
\caption{Imaginary part of dielectric function, i.e. $Im\epsilon(\omega)$ in terms of normalized frequency $\omega/t$ for different values of
spin-orbit coupling $\lambda/t$ in the absence of magnetic field for $k_{B}T/t=0.05$ under half filling condition.}
\end{flushleft}
\end{center}
\end{figure}

\begin{figure}
\begin{center}
\epsfxsize=0.8\textwidth
\includegraphics[width=12.cm]{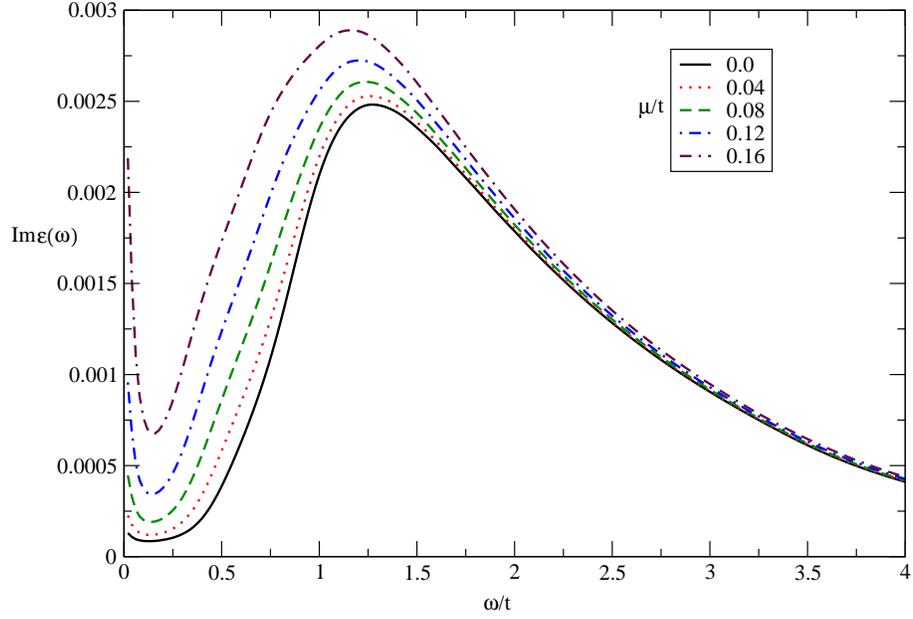}
\small
\begin{flushleft}
\caption{Imaginary part of dielectric function, i.e. $Im\epsilon(\omega)$ in terms of normalized photon frequency
 $\omega/t$ for different values of
normalized chemical potential $\mu/t$ at $\lambda/t=0.5$ in the absence of magnetic field. The temperature has been assumed to be $k_{B}T/t=0.05$}
\end{flushleft}
\end{center}
\end{figure}

\begin{figure}
\begin{center}
\epsfxsize=0.8\textwidth
\includegraphics[width=12.cm]{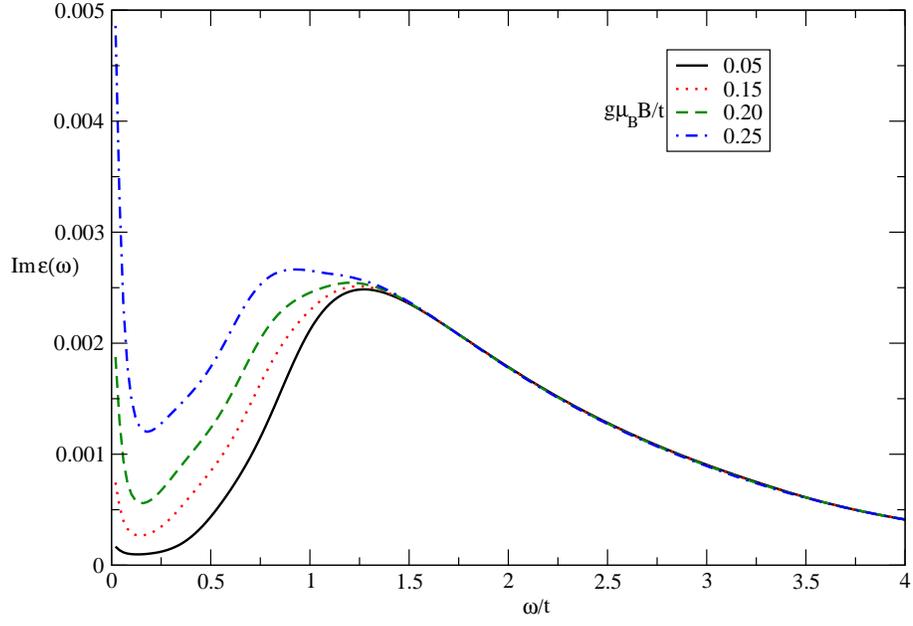}
\small
\begin{flushleft}
\caption{Imaginary part of dielectric function, i.e. $Im\epsilon(\omega)$ in terms of normalized photon frequency
 $\omega/t$ for different values of
normalized magnetic field $g\mu_{B}B/t$ at $\lambda/t=0.5$ under half filling constraint. The temperature has been assumed to be $k_{B}T/t=0.05$}
\end{flushleft}
\end{center}
\end{figure}

\section{Conflict Of ineterest Statement}
There is no conflict of ineterest statement in this manuscript.

\end{document}